\documentclass[nofootinbib,notitlepage]{revtex4-1}%

\usepackage{slashed,amsmath,amssymb}
\usepackage{graphicx}
\usepackage{dcolumn}
\usepackage{bm}
\usepackage{ dsfont }
\usepackage{xcolor}

\usepackage{color} \definecolor{darkgreen}{rgb}{0,.5,0}
\usepackage[colorlinks,filecolor=blue,citecolor=darkgreen,unicode]{hyperref}

\newcommand{\be}{\begin{equation}}\newcommand{\ee}{\end{equation}}
\newcommand{\bea}{\begin{eqnarray}}\newcommand{\eea}{\end{eqnarray}}
\def\({\left(}
\def\){\right)}
\def\[{\left[}
\def\]{\right]}
\newcommand{\pa}{\partial}
\newcommand{\la}{\lambda}
\newcommand{\om}{\omega}
\newcommand{\si}{\sigma}

\def\al{\alpha}

\def\ka{\varkappa}
\newcommand{\Ga}{\Gamma}
\newcommand{\Om}{\Omega}

\def\A{{\cal A}}

\def\C{{\cal C}}
\def\T{{\cal T}}
\def\P{{\cal P}}
\def\L{{\cal L}}
\def\M{{\cal M}}
\def\S{{\cal S}}

\def\bk{{\mathbf k}}

\def\tr{{\mathop{\rm tr}}}

\renewcommand{\phi}{\varphi}
\newcommand{\Ref}[1]{(\ref{#1})}
\newcommand{\Cite}[1]{Ref.~\onlinecite{#1}}

\newtheorem{thm}{Teorema}[section]

\begin{document}

\title{
To gap or not to gap?\\
Mass distortions and edge modes in graphene armchair nanoribbons
}

\author{C. G. Beneventano}
\affiliation{Departamento de F\'isica and Facultad de Ingenier\'ia, Universidad Nacional de La Plata,
Instituto de F\'isica de La Plata,  CONICET--Universidad Nacional de La Plata,
C.C.67, 1900 La Plata, Argentina}%
\author{I. V. Fialkovsky}%
  \email{fialkovsky.i@ufabc.edu.br}
\affiliation{%
 CMCC-Universidade Federal do ABC, Santo Andr\'e, S.P., Brazil
}%
\author{M. Nieto}%
\author{E. M. Santangelo}%
\affiliation{Departamento de F\'isica, Universidad Nacional de La Plata,
Instituto de F\'isica de La Plata,  CONICET--Universidad Nacional de La Plata,
C.C.67, 1900 La Plata, Argentina}%
\date{\today}
\begin{abstract}
We investigate, in the framework of macroscopic Dirac model, the spectrum, charge density and conductivity of metallic armchair graphene nanoribbons in presence of different mass terms. We reveal the conditions and symmetries governing the presence of edge modes in the system. Depending on the mass terms present they are exponentially localized gapless or gapped modes. The latter situation is realized, in particular, for a full Kekul\'e distortion. For this case, we calculate the mean charge and conductivity of the ribbon, and derive the traces of the presence of edge modes suitable for experimental veriﬁcation.


\end{abstract}

\pacs{Valid PACS appear here}

\maketitle


\section{Introduction}
Graphene and other allotropes of carbon are very familiar now to the scientific community, and do not require any further presentation. Since its discovery a decade ago, graphene was thoroughly studied. However, its properties are so peculiar that it is still reveling new physics. Among the most prominent examples of the effects tested in graphene one can mention unconventional Hall effect, electron Zitterbewegung, and many others \cite{CastroNeto2009,Katsnelson2012}.

Graphene nanoribbons (GNRs) came in sight as an appealing counterpart of graphene, as they allow for the opening of well modeled energy gaps \cite{CastroNeto2009,Wakabayashi2010}.
The gap creation mechanisms in graphene is a very important and demanding topic, since it paves the way for its practical utilization in nanoelectronic and optoelectronic devices \cite{Li1229}.
It was shown in numerous studies that the electronic and magnetic properties of GNRs can be tuned by the their chemical structure (edge passivation, etc), edge geometry and recently even by the engineered strain \cite{Torres2017}. The nanoribbons are usually maned, according to their edge geometry, as zig-zag and armchair ones.  While all zig-zag GNRs were shown to be metallic \cite{Brey2006a}, armchair graphene nanoribbons (AGNRs) can be divided into three distinct families depending on their width \cite{Wakabayashi1998}. Graphene nanoribbons were the subject of many a study, both recent and antique (see \cite{Wakabayashi2009,CastroNeto2009} for a review of the subject and further references). The recent technological developments made the creation and thorough investigation of atomically well-defined AGNR \cite{Kimouche2015,Wang2016} possible.

In the present paper we investigate, theoretically in the continuum limit, the metallic family of AGNRs, characterized by the presence of $N=3p+2$ ($p$ integer) atoms across the ribbon width. We study the spectrum of such nanoribbons and analyze the possibility of opening a gap and the existence, or lack thereof, of exponentially localized edge states under different distortions (i.e., types of mass terms).
Unlike commonly stated in the literature \cite{Nakada1996,Wakabayashi2010}, we show that the metallic AGNRs do, in some cases, possess edge states, whose existence and localization is governed by the mass terms present in the system.

We also study, for a full Kekul\'e distortion \cite{Hou:2006qc}, the mean charge density and the AC conductivity using the techniques of Quantum Field Theory as applied to the physics of graphene \cite{Fialkovsky2012,Fialkovsky:2016kio}. Separating the contribution of the edge modes, we identify the signatures of their presence amenable to detection in possible future experiments.


The paper is organized as follows: Section \ref{undistorted} contains the basics of undistorted AGNRs. After a brief presentation, in subsection \ref{formulation}, of the tight-binding model in the continuous limit, as well as the relevant boundary conditions and symmetries, we present its spectrum without any mass term in subsection \ref{basicspectrum}. In section \ref{masses} we present, following reference \cite{Chamon:2012vq}, the possible mass terms compatible with a Laplace-type squared Hamiltonian, and uncover the existence of a geometrical symmetry which protects the gapless edge modes. Subsection \ref{sec:diracmass} contains the determination of the spectrum and the associated eigenfunctions for a standard Dirac mass (real Kekul\'e distortion), while subsection \ref{completekekule} contains the spectral resolution in the case of a full Kekul\'e distortion. In this last case, we proceed to the investigation of the signatures of the presence of edge modes in the charge density (Subsection \ref{carga}) and in the longitudinal conductivity (Subsection \ref{conduct}). Appendix \ref{teorema} contains a theorem proving that, whenever exponential modes exist, they are, all across the ribbon, in the same subspace selected by the boundary conditions. We also prove that, among the admissible mass terms, only those respecting the aforementioned symmetry allow for the presence of gapless (perfectly conducting) edge modes. It also proves that, in the presence of a sum of two anticommuting mass terms in the Hamiltonian, one of them breaking such symmetry (the case of a full Kekul\'e distortion), the edge modes still exist but they become gapped modes. Appendices \ref{app:m5} and \ref{sec:app-pol-op} contain some details of the ingredients necessary to our calculations of mean charge and conductivity for Kekul\'e AGNRs. Finally, Appendix \ref{m3} lists the corresponding results for a different pair of anticommuting mass terms, to check the validity of a previous conjecture in reference~\onlinecite{Beneventano2014}, while extending it to the cases allowing for edge modes.

Unless stated otherwise, throughout the paper we work in ``natural'' units, $\hbar=v_F=1$. The physical units will be recovered, whenever needed, on dimensional grounds.

\section{Undistorted AGNR${\bf s}$. Symmetries and spectrum} \label{undistorted}
\subsection{The Hamiltonian formulation}\label{formulation}
The macroscopic model for the electron quasi particles in graphene is very well known, see e.g. \cite{CastroNeto2009}. In the continuum limit it is given by the Dirac Hamiltonian, which we take as in \Cite{Brey2006a}:
\begin{equation}
H_0=
    \left( \begin{array}{cc} -\sigma_1 k_1 -\sigma_2 k_2 & 0 \\
        0 & \sigma_1 k_1 -\sigma_2 k_2 \end{array}\right)
  = -  \left(k_1 \tau_3 \otimes \si_1  + k_2 \tau_0 \otimes \si_2\right).
\label{DirH}
\end{equation}
Here $\si_i$ and $\tau_i$, $i=1,2,3$ are Pauli matrices acting in the sub-lattice subspace and in the Dirac points one correspondingly, as in \cite{Beenakker2008}. $\tau_0$ is $2\times2$ identity matrix,  $k_j=-i \partial_j$, $j=1,2$.

The Hamiltonian \Ref{DirH} acts on a $4$--spinor
\be
  \psi=(\psi_A,\psi_B,-\psi_A',-\psi_B')^T,
  \label{repre}
\ee
where $\psi_{A,B}^{(')}=\psi_{A,B}^{(')}(x^1, x^2)$ are electron envelope functions corresponding to the sub-lattice $A(B)$, and the valley $K^{(')}$.

At the armchair edges of the ribbon, say at $x^1=0,W$, the spinor \Ref{repre} satisfies the following boundary conditions \cite{Brey2006a}
\be
\psi(x^1=0)=
    -\left(
      \begin{array}{cc}
        0 &  \si_0 \\
        \si_0 & 0 \\
      \end{array}
    \right)
    \psi(x^1=0),
    \ee
    \be
\psi(x^1=W)=
    -\left(
      \begin{array}{cc}
        0 & e^{i\phi} \si_0\\
        e^{-i\phi} \si_0 & 0 \\
      \end{array}
    \right)
    \psi(x^1=W)
\,,\ee
with the phase $\phi$ depending on the atomic width of the ribbon. The metallic behavior is characterized by $\phi=0$. In $\si$-$\tau$ notation the boundary conditions become
\be
\psi(x^1=0,W)=
    -\tau_1 \otimes \si_0
    \psi(x^1=0, W).
 \label{ArmBC}
\ee

We now introduce $4\times 4$ $\gamma$--matrices
\bea
\Gamma^0 &=&\( \begin{array}{cc} 0 & i\sigma_1 \\ -i\sigma_1 & 0 \end{array}\)
  =- \tau_2 \otimes \si_1,\quad
\Gamma^1 =\( \begin{array}{cc} 0 & -i\si_0 \\ -i \si_0 & 0 \end{array}\)
  =-i\tau_1 \otimes \si_0 \nonumber \\
\Gamma^2 &=&\( \begin{array}{cc} 0 & -\sigma_3 \\ \sigma_3 & 0 \end{array}\)
  = - i \tau_2 \otimes \si_3 ,\quad
\Gamma^3 =\( \begin{array}{cc} 0 & -\sigma_2 \\ \sigma_2 & 0 \end{array}\)
  = - i \tau_2 \otimes \si_2
\,,\label{Gam3}
\eea
so that the Hamiltonian \Ref{DirH} can be written as
\be
H_m
  = -  \Gamma^0 (\Gamma^1k_1+\Gamma^2k_2+ {\cal M})
    =  \Gamma^0 (i \Gamma^1 \partial_1 +i \Gamma^2 \partial_2 -  {\cal M})
    \,,\label{DirH4}
\ee
where we also introduced a generic mass term, ${\cal M}$, to be discussed in the next section, and returned to the coordinate representation explicitly.

In this $\Ga$-representation the armchair boundary conditions \Ref{ArmBC} become
the so called MIT bag \cite{MITbag}, or Berry-Mondragon \cite{BerryMondragon} ones
\begin{equation}
  \psi({x^1=0,W})=-i\Gamma^1\psi({x^1=0,W}).  \label{armbag}
\end{equation}
In the original considerations of \cite{BerryMondragon} the sign in the RHS was correlated at the opposite sides of a sample, since it was determined by the direction of the outside normal vector (see also \cite{Beneventano2014}). This is not the case for AGNRs, where such sign is the same at both boundaries.

Two discrete symmetries of the massless Hamiltonian \Ref{DirH} and of the boundary conditions \Ref{ArmBC} are spatial inversion $\P$ and time reversal $\T=T \C$,
where $\C$ is complex conjugation. In the representation \Ref{Gam3} they are given by
\be
  \P = -\tau_1 \otimes \si_1 = i \Ga^0\Ga^5,\quad
  T=  \tau_1 \otimes \si_0  =  i \Ga^1\,.
 \label{simm}
\ee
Here, $\Ga^5=i\Ga^0\Ga^1\Ga^2\Ga^3 $.

They are such that
\bea
\P H_0 (\bk)\P &=&H_0 (-\bk)\nonumber\\
T H_0^{*} (\bk)T &=&H_0 (-\bk)\,.
\label{sym1}
\eea
There are still one unitary and one antiunitary symmetry compatible with the boundary conditions (see, for instance, \cite{Wurm2012})
\bea
{\P}_0 &=& -\tau_0 \otimes \si_3 = \Ga^0\Ga^3 \nonumber\\
  \T_{0} &=& T_{0} \C,\quad {\rm{with}}\, T_0 = \tau_0 \otimes \si_2=\Ga^0 \Ga^2\,.
\eea
They are such that
\bea
{\P}_0 H_0 (\bk){\P}_0 &=&-H_0 (\bk)\nonumber\\
T_0 H_0^{*} (\bk)T_0 &=&-H_0 (\bk)\,.\label{sym2}
\eea
${\P}_0$ is the so called sublattice or chiral symmetry. As for $\T_{0}$, it is distinguished from $\T$ by the fact that $\T_{0}^2 = - 1$, while $\T ^2 =  1$. As we will see in brief, one or the other of these two last symmetries (depending on the mass term present) insure the symmetric character of the spectrum around zero.

Finally, one more unitary symmetry can be noted in the system, and it is this one which makes the behavior of the AGNRs so peculiar. We can see that
\be
  \L H_0(\bk)=H_0(\bk) \L,\qquad
    \L=\tau_1 \otimes \si_2  = -i \Ga^0 \Ga^1 \Ga^2= \Ga^5 \Ga^3
  \label{L}
\ee
and the boundary conditions \Ref{ArmBC} are also invariant under $\L$. It acts on spinor components \Ref{repre} as
\be
    \(
      \begin{array}{c}
        AK\\
        BK\\
        -AK'\\
        -BK'\\
      \end{array}
    \)
    \xrightarrow{\L}
    i \(
      \begin{array}{c}
        BK'\\
        -AK'\\
        -BK\\
        AK\\
      \end{array}
    \).
\ee
Here, we recognize a $\pi/3$ rotation of the nanoribbon: under such rotation, the armchair edge turns into itself with the sublattices shifted, while the rotation of the first Brillouin zone just interchanges the Dirac points.

\subsection{The undistorted spectrum}
\label{basicspectrum}
The spectrum of the model defined by Hamiltonian \Ref{DirH} $H_0$ with boundary conditions \Ref{armbag} is well known, see for instance \cite{Tworzydlo2006}.
It can be obtained by different methods. The eigenfunctions can be written as
\be
 \psi(x^1,x^2)=\phi(x^1) e^{i k_2 x^2}\,.
\ee

When replacing this Ansatz into $H_0  \Psi(x^1,x^2)=\Om  \Psi(x^1,x^2)$, and imposing the boundary conditions in equation (\ref{armbag}), one finds two different branches in the spectrum.
The first branch corresponds to gapped modes, with energies $\Om =\al \ka$, $\al=\pm 1$ and
\be
 \ka=\sqrt{k_1^2+k_2^2}\,,\qquad
 k_1= \pi n/W,\quad n=1,2,\ldots
\ee
Each energy level has a degeneracy of two, and the corresponding eigenfunctions are given by
\bea
\phi^{(1)}_{\al}(x^1)= N^{1}\[\sin(k_1 x^1)
    \(
    \begin{array}{c}
        -\al \ka \\
        -i k_2\\
        -\al \ka \\
        -i k_2
    \end{array}
    \)
    +i k_1 \cos(k_1 x^1)
    \(
  \begin{array}{c}
  0\\
  1\\
  0\\
  -1
  \end{array}
    \)\]\nonumber
\\
\phi^{(2)}_{\al}(x^1)= N^{2}\[\sin(k_1 x^1)
    \(
    \begin{array}{c}
        i k_2\\
        -\al \ka \\
        i k_2\\
        -\al \ka
      \end{array}
    \)
    +i k_1 \cos(k_1 x^1)
    \(
    \begin{array}{c}
	1\\
	0\\
	-1\\
	0
      \end{array}
    \)\]
\,.
\eea
Here $N^{1,2}$ are normalization constants.

The second branch corresponds to non-degenerate gapless eigenergies
\be \Om =\al  k_2,\qquad
  \al=\pm 1\,,\ee
with associated eigenfunctions
 \be \phi_{\al}^{(3)}(x^1)=N\,\(i\al,-1,-i\al,1\)^T\,.
\ee

Thus, the spectrum of the metallic AGNRs contain a gapless branch and the system shows a metallic behaviour, as the name suggests. Note that the corresponding eigenfunctions are constant across the ribbon and are eigenvectors of $\Ga^1$ with eigenvalue $i$.

\section{Possible mass terms. Presence of edge modes}
\label{masses}
As mentioned above, without any mass term metallic AGNRs are gapless and possess a perfectly conducting channel. As discussed in \cite{Gusynin2007,Chamon:2012vq}, the mass matrices which might open a gap should generally satisfy some basic anticommutation relations in order to lead to a Klein-Gordon-type squared Hamiltonian.
Namely, they must satisfy
\be
  \{\Ga^0\Ga^1,\Ga^0\M\}=\{\Ga^0\Ga^2,\Ga^0\M\}=0.
	\label{M-cond}
\ee
here $\{\cdot,\cdot\}$ is an anticommutator. However, and quite surprisingly, this condition is not always sufficient to open a gap in the considered system. Such situation is better known for two-dimensional Topological Insulators \cite{Tkachov2013}, where edge gapless modes are present, whose properties are protected by the (antiunitary) time reversal symmetry.

The mass matrices $\M$ satisfying \Ref{M-cond} are the following four:
\be
 M= m I,\qquad
 M_5= i m_5 \Ga^5, \qquad
  M_3=m_3 \Ga^3,\qquad
  M_{53}=m_{53}\Ga^5 \Ga^3\,.
\ee
where $I$ is unit $4\time4$ matrix. We recognize in the first two possible masses the real and imaginary part of a Kekul\'e distortion. In particular, $M$ is the standard Dirac mass term. The third one is a Semenoff mass, which models a staggered potential. Finally, the last one is the Haldane's mass, known to produce a quantum Hall effect without magnetic field. For a detailed study of these and others mass terms see \cite{Chamon:2012vq} and \cite{Gusynin2007}.

In the massive Hamiltonian
$$
	H_m\equiv H_0+\mathfrak{M}
$$
they correspond, respectively, to the following mass terms
\be
 \mathfrak{M}= m \Ga^0,\qquad
  \mathfrak{M}_5 =i m_5 \Ga^0\Ga^5, \qquad
  \mathfrak{M}_3 =m_3 \Ga^0\Ga^3,\qquad
  \mathfrak{M}_{53}=m_{53}\Ga^0\Ga^5 \Ga^3\,.
\ee
It can be checked which, among them, fulfill each of the aforementioned discrete symmetries (in the sense of equations (\ref{sym1}), (\ref{sym2}) and (\ref{L}), with $H_m$ instead of $H_0$). The conclusions of such analysis are shown in the following table:
\begin{center}
\begin{tabular}{|c|c|c|c|c|c|}
  \hline
   & $\P$ & $\T$ & ${\P}_0$ & ${\T}_0$ & $\L$ \\ \hline
  $\mathfrak{M}$ & No & Yes & Yes & No & Yes\\ \hline
 $\mathfrak{M}_5 $& Yes & Yes & Yes & No & No \\ \hline
  $ \mathfrak{M}_3$ & No & Yes & No & Yes & No \\ \hline
  $\mathfrak{M}_{53}$ & Yes & No & No & Yes & Yes \\
  \hline
\end{tabular}
\end{center}
We will show, in Appendix \ref{teorema} that, among these, only those mass terms preserving $\L$ admit the presence of gapless modes. Moreover, we note from our table that the symmetry of the spectrum is guaranteed in all cases, by either ${\P}_0$ or ${\T}_0$.

For the sake of our calculations in sections \ref{carga} and \ref{conduct}, we will, in what follows, study the spectrum of the modified Dirac operator
\be
	{\cal D}\equiv
	i\partial_0 - H_m.
	\label{cal D 1}
\ee

\subsection{Spectrum of the system for a standard Dirac mass term}
\label{sec:diracmass}
Let us start by considering the simplest possible mass matrix, given by a Dirac mass term
\be
\M = M = m I\,.
\ee
This type of mass is  related to the real part of a Kekul\'e distortion \cite{Hou:2006qc}.
The eigenvalue problem for the modified Dirac operator \Ref{cal D 1} is then
\be
	{\cal D}\Psi=
	(i\partial_0 - H_m)\Psi
	= \mathcal{E}\Psi
	\label{cal D}
\ee
which becomes, for $\Psi(x^0,x^1, x^2)= e^{-i k_0 x^0}\psi(x^1, x^2)$,
\be
  (\Om+H_0-   m \Ga^0)\psi = 0\,, \label{eigen}
\ee
where $\Om$ is given by
\be
	\Om  = \mathcal{E}- k_0. \label{Om-ep}
\ee

 In order to find the spectrum and eigenfunctions of equation \Ref{cal D} with armchair boundary conditios \Ref{armbag}, we write
\be
 \psi(x^1,x^2)=\phi(x^1) e^{i k_2 x^2}\,.
\ee
Then, we can treat \Ref{eigen} as a system of ordinary differential equations in the form
\be
  \frac{d \phi(x^1)}{d x^1} = {\cal A} \phi(x^1)
  \label{eq psi}
\ee
with
\be
	\A = i k_2 \Ga^1\Ga^2 + i \Om \Ga^0\Ga^1 + i m \Ga^1 \,.
	\label{cal A}
\ee
A general solution of \Ref{eq psi} is a linear combination 
\be
	\phi(x^1)  = \sum_{i=1}^4 a_i v_i e^{x^1 \la_i}\,,
\label{genSol}
\ee
where $v_i$, $\la_i$, $i=1,2,3,4$ are the eigenvectors and the corresponding eigenvalues of $\A$, i. e.
\be
	\A v_i=\la_i v_i, \qquad i=1,2,3,4.
\ee
and $a_i$ are arbitrary complex numbers.
When only a standard Dirac mass term is present it is relatively easy to find doubly degenerate eigenvalues
\be
	\la_{1,2}=\pm \sqrt{k_2^2+m^2-\Om^2},
	\label{la_12}
\ee
each of them with two associated eigenvectors
\be
	v_{i}^a=(i\Om,  \la_{i}-k_2,0,m)^T, \quad
	v_{i}^b=(\la_{i}+k_2,i\Om, m,0 )^T, \quad i=1,2.
\ee

To obtain the spectrum, i.e. the possible values of $\Omega$ and, hence of $\cal E$, we need to satisfy the boundary conditions \Ref{armbag}. They impose, on the components of the general solution \Ref{genSol}, the conditions
\begin{eqnarray}
&&\phi^1+\phi^3\big|_{x^1=0}=\phi^1+\phi^3\big|_{x^1=W}=0,\nonumber\\
&&\phi^2+\phi^4\big|_{x^1=0}=\phi^2+\phi^4\big|_{x^1=W}=0.
\label{eqsBC}
\end{eqnarray}
These four equations give us an homogeneous system of linear equations with variables $a_i$, $i=1,2,3,4$ from \Ref{genSol}. To get a nontrivial solution, the corresponding determinant must be zero. This condition, imposed on \Ref{eqsBC}, leads to
\be
  \frac{4 (k_2^2 -\Om^2 )}{m^2-\Om^2} {\rm sh}^2 (W \la_1)=0,
  \label{detBC}
\ee
which determines the possible values of $\Om$. Note that, as the previous expression is the product of two factors, there are two possibilities to obtain a vanishing determinant: either one or the other of the two factors must be zero, thus giving us two different types of solutions.

The second factor of \Ref{detBC} vanishes for
\be
	\la_1= i \pi n/W,\quad n=1,2,\ldots
\ee
From \Ref{Om-ep} and  \Ref{la_12} we find the spectrum
\be
  {\cal E}^{(1)} (\al,k_0,k_1,k_2)\equiv {\cal E}^{(2)} (\al,k_0,k_1,k_2)
  = k_0 +\al  \ka\,,\qquad
  \al=\pm 1\,.
  \label{app_mspe}
\ee
where
\be
 \ka=\sqrt{k_1^2+k_2^2+m^2}\,,\qquad
 k_1= \pi n/W,\quad n=1,2,\ldots
\ee
Each energy level has a degeneracy of two, and the corresponding eigenfunctions are given by
\bea
\phi^{(1)}_{\al}(x^1)= N^{1}\[\sin(k_1 x^1)
    \(
    \begin{array}{c}
        -\al \ka \\
        -i (k_2+m)\\
        -\al \ka \\
        -i (k_2-m)
    \end{array}
    \)
    +i k_1 \cos(k_1 x^1)
    \(
  \begin{array}{c}
  0\\
  1\\
  0\\
  -1
  \end{array}
    \)\]\nonumber
\\
\phi^{(2)}_{\al}(x^1)= N^{2}\[\sin(k_1 x^1)
    \(
    \begin{array}{c}
        i (k_2-m)\\
        -\al \ka \\
        i (k_2+m)\\
        -\al \ka
      \end{array}
    \)
    +i k_1 \cos(k_1 x^1)
    \(
    \begin{array}{c}
	1\\
	0\\
	-1\\
	0
      \end{array}
    \)\]
\,,
\label{mspe}
\eea
with $N^{1,2}$ normalization constants.

The more surprising outcome of our calculation is the presence, in metallic AGNRs, of eigenfunctions which are concentrated near one of the edges. Indeed, the second case in which the determinant \Ref{detBC} vanishes is for
\be
	k_2^2 -\Om^2=0
\ee
Then the corresponding branch of the spectrum of the problem becomes:
\be
  {\cal E}^{(3)}(\al,k_0,k_2) = k_0 + \al  k_2,\qquad
  \al=\pm 1\,. \label{ep exp m}\ee
Going back now to equations \Ref{genSol} and \Ref{eqsBC}, we find that the eigenvalues \Ref{ep exp m} have no degeneracy, and solving the linear system we obtain
\be
  \phi^{(exp)}_{\al}(x^1)=N\,e^{-mx^1}\(i\al,-1,-i\al,1\)^T,\qquad
  \al=\pm1\,.
  \label{ex_m}
\ee
So, the eigenfunction of the conducting, gapless mode of AGNR is
\be
\Psi^{(exp)}_{\al,k_0,k_2}= N e^{-i k_0 x^0} e^{i k_2 x^2}e^{-mx^1}\(i\al,-1,-i\al,1\)^T.
\ee
The constant $N$ is, again, a normalization constants.

These are edge states, exponentially localized (for non zero $m$) near one of the edges of the AGNR, due to parity breaking by $M$. Note that, at variance with the case of the ordinary (gapped) modes, there is no degeneracy for these (gapless) ones.
We see that a standard Dirac mass fails to open a gap in the system, which is consistent with our general theorem in Appendix \ref{teorema}, since $\mathfrak{M}=m\Ga^0$  commutes with $\L$, defined in \Ref{L}.

One can check that, in close resemblance to the general theory of two-dimensional topological insulators \cite{Tkachov2013} we have
\be
  {\T} \psi^{(exp)}_{\al,k_2}(x^1,x^2) = \psi^{(exp)}_{-\al,-k_2}(x),\qquad {\rm{where}} \quad
  \psi^{(exp)}_{\al,k_2}(x^1, x^2) = e^{i k_2 x^2} \phi^{(exp)}_{\al}(x^1).
\ee
Moreover, the structure of \Ref{ex_m} shows that these modes do not scatter into each other under a scalar potential, $\sim \Ga^0$
\be
\bar \phi^{(exp)}_{-\al} \Ga^ 0 \phi^{(exp)}_{\al} = 0,\qquad 
	\bar \phi \equiv \phi^\dag \Ga^ 0.
\ee
This means that the edge modes $\phi^{(exp)}$ and, thus, the conducting properties of AGNRs, are protected against scalar impurities, either of long or short range. One can also note that $\bar \phi^{(exp)}_{\al} V \phi^{(exp)}_{\al}=0$, for a potential of the form
\be
  V\sim\tau_3 \otimes \si_0 =-\Ga^5\,,
\ee
constant across sublattices, but of opposite sign at the $K$ points.

\subsection{Opening a gap}
\label{kekule}

As already said, the natural attempt to open a gap is to consider other types of mass terms \cite{Gusynin2007}. Formally speaking, one could write $\M$ as a linear combination of all four solution of equation (\ref{M-cond}) (or even of all 16 $\gamma$--matrices forming the basis), and investigate the dependence of the spectrum of the model on the coefficients of such combination. However, a guiding observation can shorten our way to the opening of a gap. One can note that the edge modes \Ref{ex_m} are actually eigenfunctions of the additional symmetry operator $\L$
\be
  \(\L +\al\) \phi^{(exp)}_{\al} = 0\,,
\ee
which suggests that the edge modes are actually protected by this symmetry. We prove that it is so in Appendix \ref{teorema}. An interesting case of a mass term breaking $\L$ is $M_5=i\,m_5 \Ga^5$, which does, indeed, lead to a gapped spectrum. So, in what follows, we will study a full Kekul\'e distortion, i.e., $\M=mI+i\,m_5 \Ga^5$.
This distortion was recently shown to be the leading mechanism of mass formation in graphene and graphene nanoribbons, \cite{Lin2017}. Thus,  such system is of evident physical interest.

Moreover, the presence of a gap will be important when performing a Pauli-Villars regularization of the integrals in our calculations of sections \ref{carga} and \ref{conduct}.

\subsection{Spectrum for the complete Kekul\'e distortion}
\label{completekekule}

As mentioned above, the most interesting and physically viable way to open a gap, while still having the localized edge modes is to consider a complete Kekul\'e distortion. So, we shall consider
\be
  \M = m I +i m_5 \Ga^5.
\ee
Following the method of Section \ref{sec:diracmass} (see Appendix \ref{app:m5} for details) we derive that in this case, the spectrum also presents a non degenerate edge branch. However, it does present a gap. Indeed, this branch of the spectrum is given by
\be
  {\cal E}^{(3)}(\al,k_0,k_2) = k_0 +\al  \sqrt{k_2^2+m_5^2},\quad \alpha =\pm 1\,.
  \label{E3_m5}
\ee
As for the the ordinary branch, which is also gapped, it is given by
\be
{\cal E}^{(1)} (\al,k_0,k_1,k_2)= {\cal E}^{(2)} (\al,k_0,k_1,k_2)
	 = k_0 +\al  \ka,\quad \alpha =\pm 1,\quad  \ka= \sqrt{k_1^2 +k_2^2+m^2+m_5^2}\,.
	   \label{E12_m5}
\ee
where $k_1=\frac{n\pi}{W},\, n\in \mathds{N}$.
The corresponding eigenfunctions are
\be
  \phi^{(exp)}_{\al}(x^1)=N e^{ -m x^1}
  \(
    - i  \al\sqrt{k_2^2+m_5^2} ,
    k_2-im_5,
    i  \al\sqrt{k_2^2+m_5^2} ,
    - k_2+im_5
   \)^T\,,
\label{phi_ex_m5}
\ee
and
\bea
\phi^{(1)}_{\al}(x^1)= N^{1}\[\sin(k_1 x^1)
    \(
    \begin{array}{c}
        \al \ka \\
        i (k_2+m+im_5)\\
        \al \ka \\
        i (k_2-m+im_5)
    \end{array}
    \)
    +i k_1 \cos(k_1 x^1)
    \(
  \begin{array}{c}
  0\\
  -1\\
  0\\
  1
  \end{array}
    \)\],\nonumber
\\
\phi^{(2)}_{\al}(x^1)= N^{2}\[\sin(k_1 x^1)
    \(
    \begin{array}{c}
        i (k_2-m-im_5)\\
        -\al \ka \\
        i (k_2+m-im_5)\\
        -\al \ka
      \end{array}
    \)
    +i k_1 \cos(k_1 x^1)
    \(
    \begin{array}{c}
	1\\
	0\\
	-1\\
	0
      \end{array}
    \)\]
    \label{phi_ord_m5}
\,.\eea

Again, $N$ and $N^{(1,2)}$ are normalization constants. For the details of this calculation, see Appendix \ref{app:m5}, where a complete set of orthonormal eigenfunctions is also constructed.

As we see, a quite interesting feature of Kekul\'e distorted AGNRs is the presence of two branches of the spectrum with two different gaps: $2m$ and $2\sqrt{m^2+m_5^2+\pi^2/W^2}$, the first one corresponding to edge modes. In light of the results of \cite{Lin2017} the search for possible localized modes in the AGNRs with Kekul\'e distortion is an appealing experimental task. In what follows we present signatures of such localization in the induced charge and longitudinal conductivity of the ribbon.

\section{Charge density and optical conductivity of Kekul\'e AGNR${\bf s}$}

We will study now some basic transport properties of Kekul\'e AGNRs: the mean charge density (proportional to the induced number of particles) and the optical (AC) conductivity, in the presence of a chemical potential $\mu$ and an impurity rate $\Gamma$. We employ the QFT approach \cite{Fialkovsky2012,Fialkovsky:2016kio}, which is essentially based on the knowledge of the propagator of the fermionic quasiparticles in nanoribbons, which we present in the first subsection.

\subsection{The fermion propagator}\label{sec:propag}
By definition, the fermion propagator $\S$ is the inverse of the Dirac operator of the system
\be
\S=\slashed{D}^{-1}, \qquad
  \slashed{D} \equiv i \pa_\mu \Ga^\mu+\M, \quad \mu=0,1,2.
\ee
supplied with appropriate boundary conditions, in our case \Ref{armbag}.
The complete set of eigenvalues and normalized eigenfunctions of the
auxiliary operator $\mathcal{D}=\Ga^0\slashed{D}$ \Ref{cal D},  permits us to write the inverse of $\slashed{D}$ as
\begin{equation}
\mathcal{S}(x,y)\equiv \mathcal{D}^{-1}(x,y)\Ga_0\,,\qquad
\mathcal{D}^{-1}(x,y)=\sum_{(a)}
    \frac{\Psi_{(a)}(x)\otimes\Psi_{(a)}^\dag(y)}
    {\mathcal{E}_{(a)}}\,.
    \label{S_pre}
\end{equation}
Here $(a)$ is a multi-index that labels a complete orthonormal set $\{\Psi_{(a)},\mathcal{E}_{(a)}\}$
of eigenspinors and corresponding eigenvalues of the operator $\mathcal{D}$ (see Appendix \ref{app:m5} for their detailed calculation).

According to the considerations above, the multi-index $(a)$ in (\ref{S_pre}) contains $\{i, \alpha,k_0,k_1,k_2\}$ with $i=1,2,3$, $\al=\pm1$, $k_1=1,2,3,\ldots$, $k_0, k_2\in\mathbb{R}$. Note that, to shorten the notation, we characterize by $i=3$ the normalized edge modes, as done in Appendix \ref{app:m5}. Then the propagator reads
\begin{equation}
\mathcal{S}(x,y)\equiv \slashed{D}^{-1}(x,y)=
    \sum_{\substack{\alpha=\pm 1}}
      \sum_{\substack{ i=1,2,3 \\ k_1\in \mathbb{N}}}
       \int dk_0\, dk_2\,
    \frac{\Psi_{\alpha,k_0,k_1,k_2}^{(i)}(x)\otimes\bar\Psi_{\alpha,k_0,k_1,k_2}^{(i)}(y)}
    {\mathcal{E}^{(i)}(\alpha,k_0,k_1,k_2)}\,,
    \label{S}
\end{equation}
where the modes are given by \Ref{ortog-m5} and \Ref{enes5}, and $\mathcal{E}$ is defined in \Ref{E3_m5} and \Ref{E12_m5} (or \Ref{app_mspe}, \Ref{ep exp m5}).
Note that (\ref{S}) is a propagator of a single species of $4\times4$-spinors.

The Fermi energy shift, $\mu$, and impurities are introduced through
\begin{equation}\label{subs}
    k_0 \to \zeta(k_0)\equiv k_0 +\mu+ i \Gamma{\rm\ sgn}k_0, \quad \Ga>0\,.
\end{equation}
The phenomenological parameter $\Ga$ introduced in this way describes, for ordinary modes, weak scalar long range impurities \cite{Prange1987}. The latter do not backscatter the edge modes, and thus cannot render their lifetime (equal to the inverse of $\Ga$) finite. We do not investigate here possible ways of scattering the edge modes. However, whatever mechanism is at work, its only possible final result is the introduction of $\Ga$.

\subsection{Charge density}
\label{carga}

In the QFT approach that we adopt in this paper the density of charge carriers, $n(x)$ is given by
\begin{equation}
    n(x)   = -i \, {\rm tr} (\Gamma^0 \mathcal{S}(x,x))
  =-i     \sum_{\substack{\alpha=\pm 1}}
      \sum_{\substack{ i=1,2,3 \\ k_1\in \mathbb{N}}}
       \int dk_0\, dk_2\,
    \frac{\( \Psi_{\alpha,k_0,k_1,k_2}^{(i)}(x)\)^\dag \Psi_{\alpha,k_0,k_1,k_2}^{(i)}(x)}
    {\mathcal{E}^{(i)}(\alpha, \zeta(k_0),k_1,k_2)}\,,
    \label{n(x)}
\end{equation}
where $ \zeta(k_0)$ is defined in \Ref{subs}. In what follows, we will calculate the contributions of edge and ordinary modes in the limit $\Gamma \to 0$, which amounts to adopting the Feynman prescription for the propagator. The charge density is $\rho(x)=-e\,n(x)$, where $-e$ is the charge of the electron.

The integral in \Ref{n(x)} is divergent. It reflects the fact that quantum field theory of Dirac fermions needs to be regularized as well known from high energy physics \cite{Peskin2015}. While it is possible to render \Ref{n(x)} finite by using a symmetry trick in the spirit of \cite{Chodos1990} (unlike the forthcoming example of polarization operator), we shall check the calculation by applying the Pauli-Villars procedure. This procedure essentially consists in verifying that all the observable effects vanish when the energy gap tends to infinity. This is the (technical) reason why we were so eager to open a gap for all modes. For the discussion of different regularizations in grapene low-frequency physics and applications of the dimensional regularization scheme see \cite{Juricic2010}.


Following the considerations of \cite{Beneventano2014, Fialkovsky:2016kio} we expect that the mean charge of the AGNR (and its conductivity considered in the next Section) will contain two contributions --- one due to the edge modes, and the other due to the ordinary ones. The latter is expected to have a universal form, independent of the peculiarities of the gapped part of spectrum. This is indeed the case as we will see from the forthcoming results. The general form of these observables is given by equation (26) of \cite{Beneventano2014} for the mean charge and by equation (38) of the same reference for the conductivity.

\subsubsection{Contribution of the edge modes}

For these modes, we explicitly calculate  the local density of particles, $n^{(exp)}(x^1 ,m,m_5)$, to reveal its localization towards one of the boundaries, a characteristic amenable for experimental verification.

In this case, the vertex in \Ref{n(x)} is given by
\be
\(\Psi_{\alpha,k_0,k_1,k_2}^{(3)}(x)\)^\dag \Psi_{\alpha,k_0,k_1,k_2}^{(3)}(x) = \frac{2 m e^{-2 m x^1}}{1-e^{-2 m W}}\,.
\ee
Then,
$$
n^{(exp)}(x^1 ,m,m_5)
  =-\frac{i}{(2\pi)^2} \frac{2 m e^{-2 m x^1}}{1-e^{-2 m W}}\sum_{\alpha=\pm1}\int dk_0\, dk_2
    \frac {1}{\mathcal{E}(\alpha,\zeta(k_0),k_2)}\,.
$$
\be
 =-   \frac{2 i m e^{-2 m x^1}}{(2\pi)^2(1-e^{-2 m W})}
    \int dk_0\, dk_2\,
    \[\frac 1{\zeta(k_0)-\sqrt{k_2^2+m_5^2}}+\frac 1{\zeta(k_0)+\sqrt{k_2^2+m_5^2}}\]\,,
  \label{n(x)_E}
\ee
where we used energy from \Ref{E3_m5} and $\zeta(k_0)$ is defined Eq. \ref{subs}.
This expression shows once again that these exponential modes are, for $m\neq 0$, edge modes. In particular, for $m>0$, the charge they carry is concentrated near the boundary $x=0$.

After changing variables, $k_0\rightarrow -k_0$, in the second term, an easy integration using the Cauchy theorem leads, in the $\Ga \to 0$ limit, to
\be
    n^{(exp)}(x^1,m,m_5) = \frac{{\rm sgn} \mu}{2\pi } \frac{2 m e^{-2 m x^1}}{1-e^{-2 m W}}  \int dk_2\,  \Theta\left(\mu^2-(k_2^2+m_5^2)\right)\,,
\ee
so that
\be
 n^{(exp)}(x^1,m,m_5)= \frac{{\rm sgn} \mu}{\pi } \frac{2 m e^{-2 m x^1}}{1-e^{-2 m W}}  \,\Theta\left(\mu^2- m_5^2\right) \sqrt{\mu^2- m_5^2}\,.
 \label{n(x)-final}
\ee

The Pauli-Villars regularization consists in subtracting from the integrand/summand in the corresponding observable quantity (equation \Ref{n(x)_E} in our case) the same expression, but with the mass gap ($m_5$) replaced by a large mass parameter $M_5$. Upon calculating the now-convergent sums and integrals, the limit $M_5\to\infty$ is to be taken. If the resulting quantity is finite, it also vanishes at $M_5\to\infty$.

In our case, equation (\ref{n(x)-final}) obviously tends to zero when $m_5\to \infty$. As a consequence, the Pauli-Villars regularized result is
\be
n^{(exp) R} = \mathop{\lim}_{M_5\to\infty} \left( n^{(exp)} (m,m_5)- n^{(exp)}(m,M_5)\right)= n^{(exp)} (m,m_5)\,.
\ee


After recovering physical units
\be
n^{(exp) R} = \frac{{\rm sgn} \mu}{\pi\hbar^2v_F^2 } \frac{2 m e^{-\frac{2 m   x^1}{\hbar v_F}}}{1-e^{-\frac{2 m   W}{\hbar v_F}}}  \,\Theta\left(\mu^2- m_5^2\right) \sqrt{\mu^2- m_5^2}\,,
\ee
where $v_F=\frac{c}{300}$ and $\mu$, $m$ and $m_5$ have dimensions of energy.

\begin{figure}
\centering \includegraphics[width=8cm]{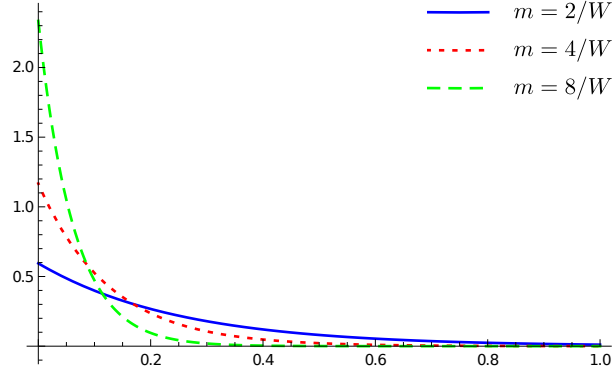}
\caption{
Charge density $n^{(exp)R}$ as a function of the relative position across the width of a ribbon, for $m=2/W, 4/W, 8/W$, $m_5=2/W$, $\mu=1.1  m_5$, in natural units.}
\label{n(x)-pict-loc}
\end{figure}

Thus, for small enough chemical potential (see \ref{charge-ord} on how small it should be), but still $|\mu|> m_5 $, we can expect the induced charge density of a metallic AGNR to be localized at one edge of the ribbon. Its profile for different values of mass to width ratio is shown in FIG. \ref{n(x)-pict-loc}.

Now, we can integrate over $x^1$ in order to obtain the mean particle number,
\begin{equation}
n(m,m_5)\equiv \frac 1W\int_0^Wdx^1\, n(x) \,,\label{j03}
\end{equation}
which, in physical units is
\be
n^{(exp)}(m,m_5)= \frac{{\rm sgn} \mu}{\pi v_F W \hbar } \,\Theta\left(\mu^2- m_5^2\right) \sqrt{\mu^2- m_5^2}\,.
\ee

\subsubsection{Contribution of the ordinary modes}\label{charge-ord}

In the case of the ordinary modes, we will calculate directly the mean number of charge carriers, which is given by

$$
   n^{(ord)}(m,m_5)=-\frac{i }{(2\pi)^2W} \sum_{\substack{\alpha=\pm 1 \\ k_1}}\int dk_0\, dk_2\,
        \frac {2}{\mathcal{E}(\alpha,\zeta(k_0),k_1,k_2)}
        $$
        \vskip-6mm
        \be
    =-\frac{2i }{(2\pi)^2W} \sum_{k_1}\int dk_0\, dk_2\,
      \[\frac 1{\zeta(k_0)-\sqrt{k_1^2+k_2^2+m_5^2+m^2}}+\frac 1{\zeta(k_0)+\sqrt{k_1^2+k_2^2+m_5^2+m^2}}\]\,,\label{j04}
\end{equation}
where $k_1=\frac{n\pi}{W},\,n\in \mathds{N}$. Note the overall factor $2$, due to the double degeneracy of each ordinary mode.
Performing the $k_0$ and $k_2$ integrals as we did in the case of the exponential modes, we arrive at
\begin{equation}
    n^{(ord)}(m,m_5)=\frac{ 2 { \rm sgn}\, \mu}{\pi W } \sum_{k_1} \Theta(\mu^2-\ka_0^2)\sqrt{\mu^2-\ka_0^2}\,,
\label{j07}
\end{equation}
where $\ka_0\equiv\ka_0(m,m_5)=\sqrt{m^2+m_5^2+k_1^2}$. As before, the contribution coming from these modes also vanishes in the $m_5\to \infty$ limit. So, once again we obtain
\be
 n^{(ord)R}= n^{(ord)}(m,m_5)\,.
 \ee

 Recovering physical units (always with $\mu$, $m$ and $m_5$ in units of energy), we have
\begin{equation}
    n^{(ord)}(m,m_5)=\frac{ 2 { \rm sgn}\, \mu}{\pi\hbar\ v_F W } \sum_{k_1} \Theta(\mu^2-\ka_0^2)\sqrt{\mu^2-\ka_0^2}\,,
\label{n-ord units}
\end{equation}
where $\ka_0\equiv\ka_0(m,m_5)=\sqrt{m^2+m_5^2+\hbar^2 v_F^2\,k_1^2},\quad k_1=\frac{n \pi}{W},\quad n=1,2\ldots$

The dependence on $k_1$ of the contribution of the ordinary modes is universal \cite{Beneventano2014,Fialkovsky:2016kio}, the only difference between different boundary conditions is residing in the allowed values of $k_1$. See, for instance, equation (26) in \Cite{Beneventano2014}, where Berry-Mondragon boundary conditions were treated (Note that, in that reference, $m$ has different units). The extra factor of two in the present calculation is due to the fact that both valleys are taken into account at once when considering the present armchair boundary conditions. Note, however, that the exponential edge modes contribute to the mean density of charge with half the contribution, as compared to ordinary ones.

As it follows from \Ref{n(x)-final} and \Ref{j07}, the contribution of the edge modes is isolated, and thus suitable for experimental verification, for $m_5< |\mu|< \sqrt{m^2+m_5^2+\hbar^2 v_F^2\,\frac{\pi^2}{W^2}}$.

\subsection{AC conductivity} \label{conduct}
Once we know the fermion propagator of the quasiparticles, we are able to use a field theoretical analog of the Kubo formula \cite{Fialkovsky2012} to calculate, again in natural units, the optical conductivity
\begin{equation}
  \label{Kubo}
  \si_{ij}(\om)=\frac{\Pi^{ij}(\om)}{i\om},\quad i,j=x,y
\end{equation}
where ${\Pi^{ij}}$ stands for the space--space components of the polarization operator
\be
 \Pi^{jk}(x,y)    =
    i  e^2  \tr\(\mathcal{S}(x,y) \gamma^j \mathcal{S}(y,x) \gamma^k\).
  \label{Pi_g}
\ee
$\mathcal{S}$ here is the fermion propagator  defined in \Ref{S}. For more details on the application of QFT techniques in describing graphene and its allotropes see reviews \cite{Fialkovsky2012,Fialkovsky:2016kio} and references therein.

\subsubsection{Polarization operator}
Combining \Ref{Pi_g} and \Ref{S} we can embrace partial translation invariance of the system (along $x^0, x^2$) to obtain
\be
\Pi^{jk}(\om,p_2;x^1,y^1)= {i C W}\int dk_0dk_2
   \,\tr\[\mathcal{S}(k_0, k_2; x^1,y^1)\Ga^j \mathcal{S}(k_0-\om, k_2-p_2; y^1,x^1)\Ga^k\]
    \label{Pi_g_Fou}
\ee
where $j,k=1,2$, $C=\frac{ e^2 }{(2 \pi)^2W}$. The Fourier representation of the propagator (\ref{S}), by using \Ref{ortog-m5} and \Ref{enes5}, becomes
\begin{equation}
\mathcal{S}(k_0, k_2; x^1,y^1) =
    \sum_{\substack{\alpha=\pm 1}}
	\sum_{\substack{ i=1,2,3 \\ k_1\in \mathbb{N}}}
    \frac{\psi_{\alpha, k_1,k_2}^{(i)}(x^1)\otimes\bar\psi_{\alpha, k_1,k_2}^{(i)}(y^1)}
    {\mathcal{E}^{(i)}(\alpha,k_0,k_1,k_2)}\,.
    \label{S_Four}
\end{equation}

The longitudinal conductivity, $\si_{22}$, averaged over the cross section of the ribbon, can be then expressed via the corresponding component of (\ref{Pi_g_Fou}) taken at $p_2=0$,
\be
  \si_{ij}(\om)=\frac{\Pi^{ij}(\om)}{i\om},\quad i,j=x,y,
\ee
\be
\Pi^{ij}(\om)=
          \frac1W
            \int_0^W dx^1\int_0^W dy^1\Pi^{ij}(\om,0;x^1,y^1).
   \label{s_pi}
\ee

Using  (\ref{S_Four})  we can rewrite \Ref{s_pi}, identifying two interaction vertices
\be \si^{ij}(\om)=C/\om\int dk_0dk_2
    \sum_{\substack{\al,k_1\\ \al',k_1'}}\sum_{n,m=1,2,3}
      \frac{V^{(i)}_{m,n'} V^{(j)}_{n',m}}
	  {{\cal E}({\alpha,k_0, k_1,k_2})\, {\cal E}({\alpha',k_0-\om, k_1', k_2})}\,,
   \label{Pi22}
\ee
where
\be
  V^{(i)}_{m,n'} = \int_0^W dy^1 \(\bar\psi^{(m)}_{\alpha, k_1,k_2}(y^1)\Ga^i\psi^{(n)}_{\alpha', k_1',k_2}(y^1)\)\,,
  \label{vertex}
\ee
and a similar expression for $ V^{(i)}_{n',m}$.

In particular, for the longitudinal conductivity, one must consider $i=j=2$. Straightforward, though cumbersome, calculations show that the contributions of ordinary and edge modes do separate in \Ref{Pi22}, since the corresponding mixed vertices vanish,
\be
  V^{(2)}_{m,3'}=V^{(2)}_{3',m} = 0, \quad m=1,2
\ee
independently of the position of the prime. Then, as expected, \Ref{Pi22} turns out to be
\be
  \si_{22}=\si_{22}^{(ord)}+\si_{22}^{(exp)}
  \label{si_summ}
\ee
where
\be
  \si_{22}^{(ord)} =  \frac{C}{\om}\int dk_0dk_2\sum_{m,n=1,2}
    \sum_{\al,k_1}\sum_{\al',k_1'}\frac1{{\cal E} {\cal E}'}
        V^{(2)}_{m,n'}
        V^{(2)}_{n',m}
        \label{si-ordi}
\ee
\be
  \si_{22}^{(exp)} =  \frac{C}{\om} \int dk_0 dk_2
    \sum_{\al,\al'} \frac1{{\cal E} {\cal E}'}
        V^{(2)}_{3,3'}
        V^{(2)}_{3',3}\,.
        \label{si-edge}
\ee

\subsubsection{Ordinary modes contribution}

It was argued in \cite{Fialkovsky:2016kio} that the contribution of the gapped, ordinary modes, has a universal form, depending on the particularities of the system only through the values of $k_1=k_1(n,m,W,\ldots)$. As we show in the Appendix \ref{sec:app-pol-op}, it is indeed the case for AGNRs, and we get the same expression as for the bag boundary conditions \cite{Beneventano2014}
\be
    \si_{22}^{(ord)R}= 2C \int dk_2\sum_{k_1}
        \(
            F +\frac{2 i \Ga}{\ka^2}+(k_1^2+m^2+m_5^2)\(G+\frac{i \pi}{\ka^3}-\frac{4 i \Ga}{ \ka^4}\)
        \)\,,
        \label{S_R}
\ee
where the summation in $k_1$ goes over all allowed values, in our case $k_1=\pi n/W, \quad n=1,2,\ldots$
The other notations in \Ref{S_R} are
\be
   F =-\frac{i \Ga}{\om(\om+2i\Ga)}
        \log\frac{(\ka+\mu)^2+\Ga^2}{(\ka+\mu)^2-(\om+i\Ga)^2}
        +(\mu\to-\mu)
        \label{FG}
\ee
$$
G= \frac{4 i \pi}{\ka\(\om^2-4\ka^2\)}
$$
$$
    +\frac{1}{\ka} \(
        \frac{\log\(\ka-i\Ga+\mu\)-\log\({\ka+\om+i\Ga+\mu}\) }{\om \(2\ka+\om\)}
            +\frac{\log\(\ka+i\Ga+\mu\)-\log\({\ka-\om-i\Ga+\mu}\)}{\om \(2\ka-\om\)}\right.
            $$
            $$
            \left.+\frac{\log\frac{\ka-\om-i\Ga+\mu}{\ka-i\Ga+\mu}}{(\om+2 i \Ga)(2\ka-\om-2 i\Ga)}
            +\frac{\log\frac{\ka+\om+i\Ga+\mu}{\ka+i\Ga+\mu}}{(\om+2 i \Ga)(2\ka+\om+2 i\Ga)}
            +(\mu\to-\mu)\)
$$

It is important to emphasize here that, for the calculation of this expression, the Pauli-Villars regularization scheme was used to deal with the divergent integral in \Ref{si-ordi}, see reference ~\onlinecite{Beneventano2014}. It amounts to subtracting from the integrand in \Ref{si-ordi} the same expression, but taken at the mass $M_5$ instead of $m_5$. After calculating the integral, the limit $M_5\to \infty$ is to be taken.

As we will see, the contribution of the edge modes will not contain any divergences. However, the regularization cannot be applied only to one of the terms in \Ref{si_summ} and we will have to perform the procedure described above also for  the edge modes contribution.

The DC limit of \Ref{S_R} is seen (again following the same steps as in\cite{Beneventano2014}) to be given, for small values of $\Ga$, by
\be
    \si_{22}^{(ord)R}(\om=0)= \frac{2\pi C}{ \Ga |\mu|}\sum_{n=1}^{\infty}
        \Theta(\mu^2-\ka_0^2)\sqrt{\mu^2-{\ka}_0^2}+\ldots\,,
        \ee
where ${\ka}_0=\sqrt{m^2+m_5^2+\(\frac{n\pi}{W}\)^2}$ and the dots stand for terms of order $\frac{1}{\sqrt{\Ga}}$ or smaller.

Note that, in the low disorder regime, these ordinary modes give a leading contribution to the DC conductivity only for $|\mu|> \sqrt{m^2+m_5^2+\(\frac{\pi}{W}\)^2}$. Recovering physical units, this is $|\mu|> \sqrt{m^2+m_5^2+\(\frac{\hbar v_F\,\pi}{W}\)^2}$.

As we will see in what follows this will give us a way to identify the contribution of the edge modes to the longitudinal conductivity.

\subsubsection{Edge modes contribution}
We focus now on the edge modes contribution to the conductivity \Ref{si-edge}
\be
  \si_{22}^{(exp)} =  \frac{C}{\om}\int dk_0 dk_2
    \sum_{\al,\al'} \frac1{{\cal E} {\cal E}'}
        V^{(2)}_{3,3'}
        V^{(2)}_{3',3}\,.
        \label{si22_edge0}
\ee
Using \Ref{ortog-m5}, \Ref{enes5} and \Ref{vertex} one gets
\be
V^{(2)}_{3,3'}=\frac{1}{2\tilde{\ka}^2}\(k_2\tilde{\ka} (\al + \al')+im_5\tilde{\ka} (\al' - \al)\)\,.
\ee

Thus,
\be
  V^{(2)}_{3,3'} V^{(2)}_{3',3} = \frac{k_2^2 (\al+\al')^2 +m_5^2 (\al-\al')^2}{2\tilde{\ka}^2} \,,
\ee
with $\tilde{\ka}=\sqrt{k_2^2+m_5^2}$. After summation over $\al,\al'=\pm1$ the integrand of \Ref{si-edge} turns into
\be
\sum_{\al,\al'} \frac{V^{(2)}_{3,3'} V^{(2)}_{3',3}}{{\cal E} {\cal E}'}
    = \frac{2(k_2^2+\zeta(k_0)\zeta(k_0-\om)-m_5 ^2)}{(\zeta^2(k_0)-\tilde{\ka}^2)(\zeta^2(k_0-\om)-\tilde{\ka}^2)}\,,
\label{integrand}
\ee
where we used the fact that
${\cal E}_+{\cal E}_-{\cal E}_+'{\cal E}_-'=\(\zeta^2(k_0)-\tilde{\varkappa}^2\) \(\zeta^2(k_0-\om)- \tilde{\varkappa}^2\right)$, here $\pm$ denotes the $\al=\pm1$. We remind the reader, that according to (\ref{subs}) $\zeta(k_0)=k_0 +\mu+ i \Gamma{\rm\ sgn}k_0$.

With impurities, the frequency integration in \Ref{si22_edge0} with \Ref{integrand} is still possible leading to
\be
\si_{22}^{(exp)} = \frac{C}{\om}\int dk_2
    \(
	H+m_5^2 (J_1+J_2)
	\)\,,
    \label{si22_edge1}
\ee
\be
  H=-\frac{2 i\Ga }{\om(\om+2 i \Ga)}
   \log \frac{(\tilde{\ka}+\mu)^2 + \Ga^2}{(\tilde{\ka}+\mu)^2 -(\om+i\Ga)^2}+(\mu \to -\mu)
\ee
\be
  J_1= \frac{2}{\tilde{\ka}}\[
    \frac{\log\frac{\tilde{\ka}+\mu-i \Ga}{\tilde{\ka}+\mu-\om-i \Ga}}{(\om+2i\Ga)(\om+2i\Ga-2 \tilde{\ka})}
	-    \frac{\log\frac{\tilde{\ka}+\mu+i \Ga}{\tilde{\ka}+\mu+\om+i \Ga}}{(\om+2i\Ga)(\om+2i\Ga+2 \tilde{\ka})}
	\]+(\mu \to -\mu)
\ee
\be
J_2= \frac{2}{\tilde{\ka}}\[
    \frac{\log\frac{\om-\tilde{\ka}+\mu+i \Ga}{\tilde{\ka}+\mu+i \Ga}
		+\log\frac{\tilde{\ka}-\om+\mu-i \Ga}{-\tilde{\ka}+\mu-i \Ga}}{\om(\om-2 \tilde{\ka})}
  -  \frac{\log\frac{\tilde{\ka}+\om+\mu+i \Ga}{-\tilde{\ka}+\mu+i \Ga}
		+\log\frac{-\tilde{\ka}-\om+\mu-i \Ga}{\tilde{\ka}+\mu-i \Ga}}{\om(\om+2 \tilde{\ka})}
\]
\ee

The integral over $k_2$ is convergent, since the integrand is $O(\tilde{\ka}^{-2})$ for $\tilde{\ka}\to\infty$. However, we are obliged to use the Pauli-Villars prescription, since it was used for the calculation of the gapped modes contribution.

For the convergent integral in $\si_{22}^{(exp)}$ this prescription simply requires to subtract the $m_5\to\infty$ limit from the original expression. We have
$$
H\mathop\sim_{m_5\to\infty} O(\tilde{\ka}^{-2}), \qquad
	J_1\mathop\sim_{m_5\to\infty} O(\tilde{\ka}^{-4}), \quad
	J_2\mathop\sim_{m_5\to\infty} -\frac{2 i \pi}{\tilde{\ka}^3} + O(\tilde{\ka}^{-4}),
$$
and these asymptotics are uniform with respect to $m_5$ and $k_2$, so that
$$
	\si_{22}^{(exp)} \mathop\rightarrow_{m_5\to\infty}
-2 i \pi m_5^2\frac{C}{\om}  \int \frac{ dk_2}{\tilde{\ka}^3} = -\frac{4 i \pi C}{\om}\,.
$$
So, the Pauli-Villars regularized result is
$$
	\si_{22}^{(exp)R}=\si_{22}^{(exp)} +\frac{4 i \pi C}{\om}\,.
$$

From this last expression, the DC limit can be evaluated, for small disorder, as in the case of the ordinary modes, to get
\be
    \si_{22}^{(exp)R}(\om=0)= \frac{2\pi C}{ \Ga |\mu|}
        \Theta(\mu^2-m_5^2)\sqrt{\mu^2-m_5^2}+\ldots \,.
        \ee
Here, as in the case of ordinary modes, and the dots  stand for terms of order $\frac{1}{\sqrt{\Ga}}$ or smaller.

Note that, in the low disorder regime, these edge modes give a leading contribution to the DC conductivity for $|\mu|> m_5$ .(same expression in physical units). So, from our previous discussion on the contribution of the ordinary modes, we conclude that the contribution of the edge modes to the DC conductivity can be easily identified by choosing  $ m_5<|\mu|< \sqrt{m^2+m_5^2+\(\frac{\hbar v_F\,\pi}{W}\)^2}$.

In the particular case of a purely real Kekul\'e distortion ($m_5=0$), the integral in \Ref{si22_edge1} is straightforward,
\be
\si_{22}^{(edge) R} (\om)_{m_5=0}
  = \frac{8 \pi \Ga C}{\om(\om+2 i \Ga)}
+ \frac{4 i \pi C}{\om}
= \frac{4 i \pi C}{\om+2 i \Ga}\,.
    \label{si22_edge2}
\ee
Note that, in this case, the edge modes become the perfectly conducting (in the pure limit, $\Ga\to 0$), gapless ones studied in Section \ref{sec:diracmass}. Their longitudinal conductivity is independent from both the value of the Dirac mass and the chemical potential.

\section{Conclusions}

To conclude, we summarize our main results. In the first place, we have shown that conducting armchair graphene nanoribbons do allow, in the presence of certain distortions, for the existence of edge (exponential) modes. This fact is opposite to the assertion that such modes are exclusive of zig-zag graphene nanoribbons. For conducting AGNRs, such modes can be gapped or ungapped, depending on the distortion (or, equivalently, on the type of mass terms added to the free Hamiltonian). More precisely, we have proven as a theorem, that these modes belong in the subspace determined by the boundary conditions, not only at the boundaries, but all across the width of the ribbon. We have also proven that such modes are gapless whenever a particular symmetry, ${\cal L}$, interpreted here as a $\frac{\pi}{3}$ rotation, is respected by the distortion.

As a particular example where gapped edge modes exist, we have treated a full (complex) Kekul\'e distortion, which can be represented as a sum of two mutually anticommuting mass terms in the Hamiltonian ($\mathfrak{M}+\mathfrak{M}_5$), which breaks the aforementioned symmetry. For this kind of distortion, of evident physical interest in view of the results in reference \cite{Lin2017}, we have determined the spectrum. Making use of such spectrum, we have calculated the density of charge carriers (equivalently, the density of charge), both local and integrated across the ribbons's width. We have also obtained the AC and DC longitudinal conductivities for disordered Kekul\'e AGNRs, making use of the methods of Quantum Field Theory adequate to the continuum limit of the tight-binding theory. In all cases, we have shown that the contributions due to the ordinary (not concentrated at the edge) modes have a universal behavior, thus confirming the previous conjecture in reference~\onlinecite{Beneventano2014}.

Even more, we have also calculated the contribution of the edge modes to the same physical properties and clearly identified the signatures of the presence of edge modes in all of them. We have determined the range of chemical potentials, i.e., gate potentials to which Kekul\'e AGNRs must be subject to in order to detect the presence of these particular modes when measuring the local density of charge carriers, the mean charge density and the DC conductivity, this last in the regime of small disorder. Checking these predictions should be a possible task in view of the present status of the research on the subject.

Finally, as an extension of the universality conjecture previously mentioned, we also show in Appendix \ref{m3}, that the results obtained for the full Kekul\'e distortion $\mathfrak{M}+\mathfrak{M}_5$ remain formally valid for a linear combination of a real Kekul\'e mass and a Semenoff or staggered potential one, $\mathfrak{M}+\mathfrak{M}_3$, another combination of distortions leading to edge modes. In fact, this is true not only when considering the contributions from the ordinary modes, but also when it comes to the contributions from the edge ones.

\section*{Acknowledgments}

The authors are indebted to D.V. Vassilevich for fruitful discussions. One of authors (I.V.F.) expresses deep gratitude to Prof.Zagorodnev for valuable remarks and hospitality in Dolgoprudniy and ackonwledges the support under the project 2012/22426-7 of Funda\c{c}\~ao de Amparo à Pesquisa do Estado de São Paulo (FAPESP). Work of C.G.B., M.N. and E.M.S. was partially supported by CONICET (PIP 688) and UNLP (Proyecto Acreditado X-748).

\appendix

\section{Existence of exponential modes or lack thereof. $\L$ protects gapless modes}
\label{teorema}

\begin{thm}{Let the eigenvalue problem
\be
H_m \psi(x^1)
  =  (-i\Gamma^0\Gamma^1\partial_1+\Gamma^0\Gamma^2k_2+{\mathfrak M}_1\, +  {\mathfrak M}_2)\psi(x^1)=\Om \psi(x^1)\,,\label{eig}\ee
with
\be
\frac{\mathbb{I}+i\Gamma^1}{2}\psi(x^1=0)=\frac{\mathbb{I}+i\Gamma^1}{2}\psi(x^1=W)=0\,,\label{bc}
\ee
where $\{{\mathfrak M}_1,\Gamma^0\Gamma^1\}=\{{\mathfrak M}_2,\Gamma^0\Gamma^1\}=\{{\mathfrak M}_1,\Gamma^0\Gamma^2\}=\{{\mathfrak M}_2,\Gamma^0\Gamma^2\}=0$ and $\{{\mathfrak M}_1,\Gamma^1\}=0, \quad [{\mathfrak M}_2,\Gamma^1]=0$. (Note: altogether this amounts to ${\mathfrak M}_1$ respecting ${\cal L}$ in equation \ref{L} and ${\mathfrak M}_2$ breaking it.)

Then:

\begin{enumerate}

\item Whenever present, the exponential solutions of equation \Ref{eig} are such that ${\psi}^{(exp)}(x^1)$ satisfies $\frac{\mathbb{I}+i\Gamma^1}{2}{\psi}^{(exp)}(x^1)=0\quad \forall x^1: 0<x^1<W$. \label{g}

\item Such exponential eigenfunctions only exist for

\begin{enumerate}

\item  ${\mathfrak M}_1 ={\mathfrak M}_2=0$, \label{caso0}

\item ${\mathfrak M}_1 =0, {\mathfrak M}_2 \neq 0$, \label{caso1}

\item ${\mathfrak M}_1 \neq 0, {\mathfrak M}_2 = 0$ and \label{caso2}

\item ${\mathfrak M}_1 \neq 0, {\mathfrak M}_2 \neq 0$, with $\{{\mathfrak M}_1,{\mathfrak M}_2\}=0$.\label{caso3}

\end{enumerate}

\item In case \ref{caso0} they are constant rather than exponential modes. Moreover, they are gapless. In case \ref{caso1}, they are gapped constant modes. In case \ref{caso2} they are gapless edge modes. Finally, in case \ref{caso3}, they are gapped edge modes.

\end{enumerate}}\end{thm}

\textit{Proof}
From equation \Ref{eig} we have
\be
(-i\Gamma^0\Gamma^1\partial_1+\Gamma^0\Gamma^2k_2+ {\mathfrak M}_1+{\mathfrak M}_2)\psi(x^1)=\Om \psi(x^1)\,.\label{e1}\ee

Multiplying this last with $i\Gamma^1$ and using the (anti) commutation properties of the mass terms, we obtain
\be
(i\Gamma^0\Gamma^1\partial_1+\Gamma^0\Gamma^2k_2- {\mathfrak M}_1+{\mathfrak M}_2)i\Gamma^1\psi(x^1)=\Om i\Gamma^1\psi(x^1)\,.\label{e2}\ee

By adding and subtracting equations \Ref{e1} and \Ref{e2}, we get
\be
(-i\Gamma^0\Gamma^1\partial_1+ {\mathfrak M}_1)\frac{\mathbb{I}-i\Gamma^1}{2}\psi(x^1)+(\Gamma^0\Gamma^2k_2+{\mathfrak M}_2-\Om)\frac{\mathbb{I}+i\Gamma^1}{2}\psi(x^1)=0\,,\label{e12}\ee
\be
(-i\Gamma^0\Gamma^1\partial_1+ {\mathfrak M}_1)\frac{\mathbb{I}+i\Gamma^1}{2}\psi(x^1)+(\Gamma^0\Gamma^2k_2+{\mathfrak M}_2-\Om)\frac{\mathbb{I}-i\Gamma^1}{2}\psi(x^1)=0\,.\label{e22}\ee

Suppose an exponential solution to this set of equations does exist. Its general expression is ${\psi}^{(exp)}(x^1)=e^{Ax^1}v$, with $A$ an as yet undetermined, constant, $4\times 4$ matrix and $v$ a constant spinor. From the boundary condition \Ref{bc}, at $x^1=0$ one has $\frac{\mathbb{I}+i\Gamma^1}{2}v=0$ or, equivalently,$\frac{\mathbb{I}-i\Gamma^1}{2}v=v$ . Now, the boundary condition at $x^1=W$ reads $\frac{\mathbb{I}+i\Gamma^1}{2}e^{Ax^1}\frac{\mathbb{I}-i\Gamma^1}{2}v$. This can only be satisfied for a matrix $A$ such that $[A,\Gamma^1]=0$. As a consequence, $\frac{\mathbb{I}+i\Gamma^1}{2} {\psi}^{(exp)}(x^1)=0\quad \forall x^1: 0<x^1<W$, which proves \ref{g}.

For such eigenfunction, equation \Ref{e12} reads
\be
(-i\Gamma^0\Gamma^1\partial_1+ {\mathfrak M}_1)\frac{\mathbb{I}-i\Gamma^1}{2}\psi^{(exp)}(x^1)=(-i\Gamma^0\Gamma^1\partial_1+ {\mathfrak M}_1)\psi^{(exp)}(x^1)0\,. \ee

So,
\be {\psi}^{(exp)}(x^1)=e^{i\Gamma^0\Gamma^1{\mathfrak M}_1 x^1}v, \label{psi}\ee
which determines the matrix $A$ in the argument of the exponential. Now, from equation \Ref{e22}, whenever they exist, the exponential modes must also satisfy
\be
(\Gamma^0\Gamma^2k_2+{\mathfrak M}_2-\Om)\psi(x^1)=0\,,
\ee

This last equation shows that, for these modes, the corresponding eigenvalue $\Om$ is given by $\Om =\pm \sqrt{k_2^2 + m_2^2}$. From this fact and equation \Ref{psi}, the validity of \ref{caso0} to \ref{caso3} are easy to derive.

\section{Eigenfunctions for $\M=M+M_5$}\label{app:m5}

The eigenvalue problem for the modified Dirac operator \Ref{cal D 1} is
\be
	{\cal D}\Psi=
	(i\partial_0 - H_m)\Psi
	= \mathcal{E}\Psi
	\label{app-cal D}
\ee
which becomes, for $\Psi(x^0,x^1, x^2)= e^{-i k_0 x^0}\psi(x^1, x^2)$,
\be
  (\Om+H_0-  \Ga^0(mI+i\, \Ga^5 m_5))\psi = 0\,,
  \label{app-eigen}
\ee
where $\Om$ is given by
\be
	\Om  = \mathcal{E}- k_0 .
\ee

 In order to find the spectrum and eigenfunctions of equation \Ref{app-cal D} with armchair boundary conditions \Ref{armbag}, we write
\be
 \psi(x^1,x^2)=\phi(x^1) e^{i k_2 x^2}\,.
\ee
Then, we can treat \Ref{app-eigen} as a system of ordinary differential equations in the form
\be
  \frac{d \phi(x^1)}{d x^1} = {\cal A} \phi(x^1)
\label{eq psi5}
\ee
with
\be
	\A = i k_2 \Ga^1\Ga^2 + i \Om \Ga^0\Ga^1 + i m \Ga^1-m_5 \Ga^1\Ga^5\,.
	\label{cal A5}
\ee
A general solution of \Ref{eq psi5} is a linear combination 
\be
	\phi(x^1)  = \sum_{i=1}^4 a_i v_i e^{x^1 \la_i}\,,
\label{genSol5}
\ee
where $v_i$, $\la_i$, $i=1,2,3,4$ are the eigenvectors and the corresponding eigenvalues of $\A$, i. e.
\be
	\A v_i=\la_i v_i, \qquad i=1,2,3,4.
\ee
and $a_i$ are arbitrary complex numbers.
The matrix $\A$ has two doubly degenerate eigenvalues
\be
	\la_{1,2}=\pm \sqrt{k_2^2+m^2+m_5^2-\Om^2}\,,
	\label{app-la1}
\ee
each of them with two associated eigenvectors
\be
	v_{i}^a=(i\Om,  \la_{i}-k_2,0,m+i\,m_5)^T, \,
	v_{i}^b=(\la_{i}+k_2,i\Om, m+i\,m_5,0 )^T, \quad i=1,2.
\ee

To obtain the spectrum, i.e. the possible values of $\Omega$ and, hence of $\cal E$, we need to satisfy the boundary conditions \Ref{armbag}. They impose, on the components of the general solution \Ref{genSol5}, the conditions
\begin{eqnarray}
&&\phi^1+\phi^3\big|_{x^1=0}=\phi^1+\phi^3\big|_{x^1=W}=0,\nonumber\\
&&\phi^2+\phi^4\big|_{x^1=0}=\phi^2+\phi^4\big|_{x^1=W}=0.
\label{eqsBC5}
\end{eqnarray}
These four equations give us an homogeneous system of linear equations with variables $a_i$, $i=1,2,3,4$ from \Ref{genSol5}. To get a nontrivial solution, the corresponding determinant must be zero. This condition, imposed on \Ref{eqsBC5}, leads to
\be
 (m^2-\la_1^2) {\rm sh}^2 (W \la_1)=0,
  \label{detBC5}
\ee
which determines the possible values of $\Om$ as function of $\la_1$, \Ref{app-la1}. Note that, as the previous expression is the product of two factors, there are two possibilities to obtain a vanishing determinant: Either one or the other of the two factors must be zero, thus giving us two different types of solutions.

The second factor of \Ref{detBC5} vanishes for
\be
	\la_1= i \pi n/W,\quad n=1,2,\ldots
\ee
From \Ref{Om-ep} we find the spectrum
\be
  {\cal E} ^ {(1)}(\al,k_0,k_1,k_2)\equiv
    {\cal E} ^ {(2)}(\al,k_0,k_1,k_2)= k_0 +\al  \ka\,,\qquad
  \al=\pm 1\,.
  \label{app_mspe}
\ee
where
\be
 \ka=\sqrt{k_1^2+k_2^2+m^2+m_5^2}\,,\qquad
 k_1= \pi n/W,\quad n=1,2,\ldots
\ee
Each energy level has a degeneracy of two, and the corresponding eigenfunctions are given by
\bea
\phi^{(1)}_{\al}(x^1)= N^{1}\[\sin(k_1 x^1)
    \(
    \begin{array}{c}
        \al \ka \\
        i (k_2+m+im_5)\\
        \al \ka \\
        i (k_2-m+im_5)
    \end{array}
    \)
    +i k_1 \cos(k_1 x^1)
    \(
  \begin{array}{c}
  0\\
  -1\\
  0\\
  1
  \end{array}
    \)\],\nonumber
\\
\phi^{(2)}_{\al}(x^1)= N^{2}\[\sin(k_1 x^1)
    \(
    \begin{array}{c}
        i (k_2-m-im_5)\\
        -\al \ka \\
        i (k_2+m-im_5)\\
        -\al \ka
      \end{array}
    \)
    +i k_1 \cos(k_1 x^1)
    \(
    \begin{array}{c}
	1\\
	0\\
	-1\\
	0
      \end{array}
    \)\]
\,.\label{ord-m5}\eea

The other possibility to obtain a vanishing determinant \Ref{detBC5} is
\be
	k_2^2+m_5^2 -\Om^2=0\,.
\ee
Then, the corresponding branch of the spectrum of the problem becomes:
\be
  {\cal E}^ {(3)}(\al,k_0,k_2) = k_0 + \al  \sqrt{k_2^2+m_5^2},\qquad
  \al=\pm 1\,. \label{ep exp m5}\ee
Going back now to equations \Ref{genSol5} and \Ref{eqsBC5}, we find that the eigenvalues \Ref{ep exp m5} have no degeneracy, and solving the linear system we obtain for the gapped edge modes
\be
  \phi^{(exp)}_{\al}(x^1)=N e^{ -m x^1}
  \(
    - i  \al\sqrt{k_2^2+m_5^2} ,
    k_2-im_5,
    i  \al\sqrt{k_2^2+m_5^2} ,
    - k_2+im_5
   \)^T\,.
\label{ex_m5}\ee

We see that the addition of $m_5$ opened a gap in the system, in agreement with our general theorem in Appendix \ref{teorema}.

The constants $N^{1}$, $N^{2}$ and $N^{3}$ appearing in equations \Ref{ex_m5} and \Ref{ord-m5} are normalization constants. From these eigenfunctions, a complete set of orthonormal modes can be obtained. They are given by
\bea
&&\phi^{(1)}_{\al,k_1,k_2}(x^1)= N^{1}\[\sin(k_1 x^1)
    \(
    \begin{array}{c}
        \al \ka \\
        i (k_2+m+im_5)\\
        \al \ka \\
        i (k_2-m+im_5)
    \end{array}
    \)
    +i k_1 \cos(k_1 x^1)
    \(
  \begin{array}{c}
  0\\
  -1\\
  0\\
  1
  \end{array}
    \)\],\nonumber
\\
&&\phi^{(2)}_{\al,k_1,k_2}(x^1)= N^{2}\[\sin(k_1 x^1)
    \(
    \begin{array}{c}
        -i m\\
        \frac{-(k_1^2+m^2)+m(k_2-i\,m_5)}{\al \ka} \\
        i m\\
        \frac{-(k_1^2+m^2)-m(k_2-i\,m_5)}{\al \ka}
      \end{array}
    \)
    + \cos(k_1 x^1)
    \(
    \begin{array}{c}
	i k_1\\
	\frac{-k_1(k_2-i\,m_5)}{\al \ka} \\
	-ik_1\\
	\frac{k_1(k_2-i\,m_5)}{\al \ka}
      \end{array}
    \)\]\,\nonumber\\
&&\phi^{(3)}_{\al,k_2}(x^1)=N^{3}e^{ -m x^1} \(
    \begin{array}{c}
        - i  \al\sqrt{k_2^2+m_5^2}\\
        k_2-im_5 \\
         i  \al\sqrt{k_2^2+m_5^2} \\
         - k_2+im_5
      \end{array}\)
\,.\label{ortog-m5}\eea

The normalization factor are as follows:
\bea
&& N^{1}=\( 2\ka^2 W \)^{-\frac12},\qquad N^{2}=\(2(k_1^2+m^2) W \)^{-\frac12}\nonumber \\
&& N^{3}=\(\frac{\ka^2+k_2^2+m_5^2}{m}(1- e^{-2mW}) \)^{-\frac12}\,.\label{enes5}
\eea

\section{Polarization operator for the ordinary Kekul\'e modes}\label{sec:app-pol-op}

In this appendix, we give details of the calculation leading to equation (\ref{S_R}). As stated in the body of the paper,
\be
\si_{22}^{({\rm ord})}(\om)=\frac{C}{\om}\int dk_0dk_2
    \sum_{\al,\al', k_1, k_1'}\frac{\delta_{k_1 k_1'}}{{\cal E} {\cal E}'}
        \( V^{(2)}_{1,1'}V^{(2)}_{1',1}+
	  V^{(2)}_{1',2} V^{(2)}_{2, 1'}+
	  V^{(2)}_{1,2'}V^{(2)}_{2', 1}+
	  V^{(2)}_{2,2'}V^{(2)}_{2',2}
	\).
	\label{Pi22-tmp5}
	\ee
The required vertices, obtained by using equations (\ref{ortog-m5}) and (\ref{enes5}), are given by
\be
   V^{(2)}_{1,1'}= \frac{1}{2\ka^2}\(k_2\ka (\al + \al')-im_5\ka (\al' - \al)\)\,,
    \ee
    \be
  V^{(2)}_{1,2'}=V^{(2)}_{2,1'}
      = \frac{i}{2\ka}\sqrt{k_1^2 +m^2}(\al - \al')\,,
\ee
\be
  V^{(2)}_{2,2'}= \frac{1}{2\ka^2}\(k_2\ka (\al + \al')+im_5\ka (\al' - \al)\)\,.
\ee
After summation over $\al,\al'=\pm 1$ and taking into account
that ${\cal E}_+{\cal E}_-{\cal E}_+'{\cal E}_-'=\(\zeta^2(k_0)-\ka^2\) \(\zeta^2(k_0-\om)-\ka^2\)$ (here, $\pm$ denotes $\al=\pm1$)
\Ref{Pi22-tmp5} becomes
\be
\si_{22}^{({\rm ord})}(\om)  = 4\frac{C}{\om}\int dk_0dk_2
    \sum_{k_1}\frac{\zeta(k_0)\zeta(k_0-\om)+(k_2^2-k_1^2-m^2-m_5^2)}{\(\zeta^2(k_0)-\ka^2\) \(\zeta^2(k_0-\om) -\ka^2\)}\,,
\ee
which has the same form as (34)  in Ref.~\onlinecite{Beneventano2014}, except for an overall factor of $2$, which is due to the fact that here we are considering $4\times4$ fermions. Obviously, the allowed values of $k_1$ also differ, due too the different boundary conditions.

This proves the point conjectured in Ref.~\onlinecite{Beneventano2014} that the final expression for AC conductivity (39) of Ref.~\onlinecite{Beneventano2014} (along with those for carriers density and quantum capacitance) for a set of gapped modes in a nanoribbon is universal whenever the boundary conditions do not mix transversal and longitudinal momenta.  It incorporates the properties of the nanoribbon only through the allowed values of the transversal quantized momentum $k_1=k_1(m,W,\ldots)$. So, following the same steps as in that reference, one gets equation (\ref{S_R}).

\section{Eigenfunctions, charge and conductivity for $\M=M+M_3$}\label{m3}

In this appendix we list the eigenfunctions and associated spectrum of the AGNRs, i.e. the eigenfunctions of $H_m$ (equation \Ref{DirH4}), for a mass term $\M=m I +m_3 \Ga^3$.

Following a similar procedure as the one we used in Appendix \ref{app:m5}, we find, for the ordinary eigenfunctions, $\Psi^{(i)}_{\alpha,k_0,k_1,k_2}(x^0,x^1,x^2)= e^{-ik_0 x^0+i k_2 x^2}\phi^{(i)}_{\alpha,k_1,k_2}(x^1)$, $i=1,2$, with
\be
\phi_{\al}^{(1)}= \sin(k_1 x^1)
    \(
    \begin{array}{c}
        m_3-\al \ka \\
        -i (k_2+m)\\
        m_3-\al \ka \\
        -i (k_2-m)
    \end{array}
    \)
    +i k_1 \cos(k_1 x^1)
    \(
  \begin{array}{c}
  0\\
  1\\
  0\\
  -1
  \end{array}
    \)
\ee
\be
\phi_{\al}^{(2)}= \sin(k_1 x^1)
    \(
    \begin{array}{c}
        i (k_2-m)\\
        -(m_3+\al \ka )\\
        i (k_2+m)\\
        -(m_3+\al \ka )
      \end{array}
    \)
    +i k_1 \cos(k_1 x^1)
    \(
    \begin{array}{c}
	1\\
	0\\
	-1\\
	0
      \end{array}
    \)\,,
\ee
associated to
\be
  {\cal E}^{(1)} (\al,k_0,k_1,k_2)
  	\equiv {\cal E}^{(2)} (\al,k_0,k_1,k_2)= k_0 +\al  \ka\,,\qquad
  \al=\pm 1\,.
\ee
where
\be
 \ka=\sqrt{k_1^2+k_2^2+m^2+m_3^2}\,,\qquad
 k_1= \pi n/W,\quad n=1,2,\ldots
\ee

As expected from our theorem in Section \ref{teorema}, there is also a nondegenerate edge mode, given by
\be
  \Psi^{(3)}_{\alpha,k_0,k_2}
      = N e^{-ik_0 x^0+i k_2 x^2} e^{ -m x^1}\(i  ( {m_3-\al\sqrt{k_2^2+m_3^2}}) , {k_2},
	-i ({m_3-\al\sqrt{k_2^2+m_3^2}}), -{k_2}\)^T\,,
\ee
associated to
\be
  {\cal E} ^{(3)}(\al,k_0,k_2)= k_0 +\al \sqrt{k_2^2+m_3^2}\,,\qquad
 k_1= \pi n/W,\quad n=1,2,\ldots,\qquad
  \al=\pm 1\,.
\ee

From a comparison of the spectra it is obvious, that upon transforming these eigenfunctions into a complete orthonormal system $\{\phi_{\al}^{(1)}\phi_{\al}^{(2)}\phi_{\al}^{(3)}\}$,  we obtain  the density of carriers and the mean charge density result formally identical to the ones in Section \ref{carga}, with the only replacement $m_5 \rightarrow m_3$. The validity of this assertion is less obvious in the case of the longitudinal conductivity. Again, it is easy to see that the contributions of ordinary and edge modes decouple. The nonvanishing vertices are given by
\be
   V^{(2)}_{1,1'}=V^{(2)}_{1',1}= \delta_{k_1 k_1'}
    \frac{k_2 (\al \al'\ka-2 m_3)}{2 \ka \sqrt{(\ka-\al m_3)(\ka-\al' m_3)}}\,,
    \ee
    \be
  V^{(2)}_{1',2}=V^{(2)}_{2,1'}
      = \delta_{k_1 k_1'}
      \frac{k_2 ((1-\al \al')\ka + (\al'-\al) m_3)}
	{2 |k_2| (\al \ka-m_3)}\sqrt{\frac{k_1^2+m^2}{(\ka-\al m_3)(\ka-\al' m_3)} }\,,
\ee
\be
  V^{(2)}_{2,2'}= \delta_{k_1 k_1'}
    \frac{k_2 (\al \al'\ka-2 m_3)}{2 \ka \sqrt{(\ka-\al m_3)(\ka-\al' m_3)}}\,,
\ee
\be
V^{(2)}_{3,3'}=V^{(2)}_{3,3'}=\frac{2k_2 m_3+(\al + \al')k_2\sqrt{k_2^2 +m_3^2}}{2\((k_2^2+m_3^2+\al m_3\sqrt{k_2^2 +m_3^2})(k_2^2+m_3^2+\al' m_3\sqrt{k_2^2 +m_3^2})\)^{\frac12}}\,.
\ee

Although both the eigenfunctions and relevant vertices are, in this case, different from the ones obtained for a full Kekul\'e distortion, the final expression for the longitudinal conductivity is formally identical to the one obtained in the last case, with $m_5 \rightarrow m_3$, again in agreement with the conjecture in Ref.~\onlinecite{Beneventano2014}, here extended to the cases where there are edge modes.

\bibliography{ribbons09}
\bibliographystyle{apsrev4-1}

\end{document}